\documentclass[aps,prb,final,twocolumn]{revtex4-1}

\usepackage{graphicx}
\usepackage{epsfig}
\usepackage{natbib}

\begin{document}


\title{Stability of skyrmion textures and the role of thermal fluctuations 
in cubic helimagnets: a new intermediate phase at low temperature}


\author{Victor Laliena}
\email[]{laliena@unizar.es}

\author{Javier Campo}
\email[]{javier.campo@csic.es}
\affiliation{Instituto de Ciencia de Materiales de Arag\'on 
(CSIC -- Universidad de Zaragoza)}
\affiliation{\mbox{Departamento de F\'{\i}sica de la Materia Condensada, 
Universidad de Zaragoza,} \\
C/Pedro Cerbuna 12, 50009 Zaragoza, Spain}



\date{September 27, 2017}

\begin{abstract}
The stability of the four known stationary points of the cubic helimagnet energy functional:
the ferromagnetic state, the conical helix, the conical helicoid, 
and the skyrmion lattice, is studied by solving the corresponding spectral problem. 
The only stable points are the ferromagnetic state at high magnetic field and the 
conical helix at low field, and there is no metastable state. 
Thermal fluctuations around the stationary point, included to quadratic order in the 
saddle point expansion, destabilize the conical helix in a region where the ferromagnetic
state is unstable. Thus, a new intermediate phase appears which,
in a region of the phase diagram, is a skyrmion lattice stabilized by thermal fluctuations. 
The skyrmion lattice lost the stability by lowering temperature, and a new intermediate phase 
of unknown type, presumably with three dimensional modulations, appears 
in the lower temperature region.
\end{abstract}

\pacs{111222-k}

\maketitle


\section{Introduction}

Skyrmion textures that have been discovered in cubic magnets without inversion 
symmetry \cite{Muehlbauer09,Pappas09,Munzer10,Yu10} were theoretically predicted
long ago \cite{Bogdanov94a}.
However, most of the theoretical studies of cubic helimagnets have been devoted to analyze 
the stationary points of the appropriate Landau functional. 
The stability of the more complex stationary points, the skyrmion lattices (SKL), 
has been studied only in a limited way, mostly concerning with radial or elliptic
stability \cite{Bogdanov94b,Bogdanov99}.
Hence, solitonic configurations that are considered 
stable or metastable may actually be unstable due to some mode
that is non homogeneous along the magnetic field direction.
Closely related to the stability analysis is the idea, put forward in 
Ref.~\onlinecite{Muehlbauer09}, 
that some metastable state may become the equilibrium state by virtue of the thermal 
fluctuations. This idea is supported by Monte Carlo simulations \cite{Buhrandt13}. 

This work addresses these two related problems: the stability of stationary points
and the effect of thermal fluctuations at relatively low temperatures.
As a result, the phase diagram at low $T$ is determined.

The paper is organized as follows.
The next two sections are devoted to describe the model and the saddle point expansion used to
solve it. Then, the stability and the free energy including thermal fluctuations at gaussian level
are analyzed for the four known stationary points: the FM state, the conical helix (CH), 
the conical helicoid \cite{Laliena16a}, and the SKL, which is treated in the circular cell 
approximation \cite{Bogdanov94a}. 
To conclude, the resulting phase diagram is analyzed and a brief discussion of the results
is given.

\section{Model}

Consider a classical spin system on a cubic lattice with parameter $a$, whose dynamics
is governed by the Hamiltonian
\begin{equation}
\mathcal{H} = -J\sum_{\vec{r},\hat{\mu}}\vec{S}_{\vec{r}}\cdot\vec{S}_{\vec{r}+\hat{\mu}}
-D\sum_{\vec{r},\hat{\mu}}\hat{\mu}\cdot(\vec{S}_{\vec{r}}\times\vec{S}_{\vec{r}+\hat{\mu}})
-\vec{B}\cdot\sum_{\vec{r}}\vec{S}_{\vec{r}}, \label{eq:Hamil}
\end{equation}
where $J>0$ and $D>0$ are the strength of the Heisenberg exchange and Dzyaloshinkii-Moriya
coupling constants,
and $\vec{B}$ is proportional to the applied magnetic field. 
The vector $\vec{r}$ labels the lattice sites and
the unit vector $\hat{\mu}$ runs over the right-handed orthonormal triad
$\{\hat{x},\hat{y},\hat{z}\}$.
For smooth spin configurations we may take the continuum limit
\begin{equation}
\vec{S}_{\vec{r}+\hat{\mu}}=\vec{S}_{\vec{r}}+a(\hat{\mu}\cdot\nabla)\vec{S}_{\vec{r}} + O(a^2).
\label{eq:contLim}
\end{equation}
Writing the spin variable in terms of a unit vector field $\hat{n}(\vec{r})$ as
$\vec{S}_{\vec{r}}=S\hat{n}(\vec{r})$, where $S$ is the spin modulus, extracting a global factor
$JS^2/a$, plugging Eq.~(\ref{eq:contLim}) into Eq.~(\ref{eq:Hamil}), and ignoring irrelevant 
constant terms, we have
\begin{equation}
\mathcal{H} = \frac{JS^2}{a}\sum_{\vec{r}}a^3 \left(
\frac{1}{2}\partial_i\hat{n}\cdot\partial_i\hat{n}
+q_0\hat{n}\cdot\nabla\times\hat{n}-\vec{h}\cdot\hat{n}\right), \qquad
\end{equation}
where $q_0=D/Ja$ and $\vec{h} = \vec{B}/JSa^2$. Replacing
$\sum_{\vec{r}}a^3$ by $\int d^3x$, we get $\mathcal{H}=\epsilon_0\mathcal{W}$,
with $\epsilon_0=JS^2/q_0a$ and 
\begin{equation}
\mathcal{W} = q_0 \int d^3x \left(\frac{1}{2}\partial_i\hat{n}\cdot\partial_i\hat{n}
+q_0\hat{n}\cdot\nabla\times\hat{n}-q_0^2\vec{h}\cdot\hat{n}\right). \qquad \label{eq:W}
\end{equation}
In the above expression $\partial_i=\partial/\partial x_i$, and repeated indices 
are understood to be summed throughout this paper.
The constant $q_0$ has the dimensions of inverse length and sets the scale for the spatial 
modulation of the ground state, $L_0=2\pi/q_0$.
The statistical properties are given by the partition function, 
\begin{equation}
\mathcal{Z} = \int [d^2\hat{n}] \exp[-c_0\mathcal{W}],
\end{equation}
where $c_0=T_0/T$, with $T_0=\epsilon_0/k_\mathrm{B}$.

\section{Saddle point expansion}

For systems that can be described by a continuum model $1/q_0a$ is a large number
(it is about 7 in the typical cubic helimagnet MnSi), and thus 
$c_0$ is a large number provided the temperature is not too high. In that case the partition 
function can be obtained by the saddle point expansion, as follows. 
Let $\hat{n}_0$ be a stationary point, that is, a solution of the Euler--Lagrange equations, 
$\delta\mathcal{W}/\delta\hat{n}=0$, and write $\hat{n}$ in terms of two fields $\xi_\alpha$
($\alpha=1,2$) as
\begin{equation}
\hat{n} = \sqrt{1-\xi^2}\hat{n}_0+\xi_\alpha\hat{e}_\alpha, 
\end{equation}
where the three unit vectors $\{\hat{e}_1,\hat{e}_2,\hat{n}_0\}$ form a right-handed orthonormal 
triad. They can be parametrized in terms of two angles $\theta$ and $\psi$ as 
\begin{eqnarray}
\hat{e}_1 &=& (\cos\theta\cos\psi,\cos\theta\sin\psi,-\sin\theta), \\
\hat{e}_2 &=& (-\sin\psi,\cos\psi,0), \\
\hat{n}_0 &=& (\sin\theta\cos\psi,\sin\theta\sin\psi,\cos\theta).
\end{eqnarray}
Let us expand $\mathcal{W}$ in powers of $\xi_\alpha$ up to quadratic order:
\begin{equation}
\mathcal{W} = \mathcal{W}(\hat{n}_0) + q_0 \int d^3x \xi_\alpha K_{\alpha\beta}\xi_\beta
+ O(\xi^3), \label{eq:Wexp}
\end{equation}
with
\begin{eqnarray}
K_{\alpha\beta} &=& -[\nabla^2 + 2w(\hat{n}_0) + q_0^2\vec{h}\cdot\hat{n}_0]\delta_{\alpha\beta}
+\partial_i\hat{e}_\alpha\cdot\partial_i\hat{e}_\beta 
\nonumber \\
&+&q_0(\hat{e}_\alpha\cdot\nabla\times\hat{e}_\beta+\hat{e}_\beta\cdot\nabla\times\hat{e}_\alpha)
\nonumber \\
&-& (2\vec{G}\cdot\nabla+\nabla\cdot\vec{G})\epsilon_{\alpha\beta},
\label{eq:K}
\end{eqnarray}
where $G_i=\hat{e}_1\cdot\partial_i\hat{e}_2+q_0\hat{n}_{0i}$,
$\epsilon_{\alpha\beta}$ is the two dimensional antisymmetric unit tensor, and $w(\hat{n})$ 
the integrand of Eq.~(\ref{eq:W}). The term $\nabla\cdot\vec{G}\epsilon_{\alpha\beta}$ does not 
contribute to the quadratic form entering Eq. (10), due to its antisymmetry in $\alpha\beta$.
However, it has to be introduced in order to make the operator $K$ symmetric.
The linear term in Eq.~(\ref{eq:Wexp}) vanishes by virtue of the Euler-Lagrange equatios.

The fluctuation operator $K_{\alpha\beta}$ is a symmetric differential operator that is positive
definite if $\hat{n}_0$ is a local minimum of $\mathcal{W}$. In this case the free energy,
$\mathcal{F}=-(1/c_0)\ln\mathcal{Z}$, can be
obtained from the saddle point method \cite{ZinnJustin97}, which is an asymptotic expansion 
in powers of $1/c_0$ that to lowest order, ignoring some irrelevant constants, gives
\begin{equation}
\mathcal{F} = \mathcal{W}(\hat{n}_0) + (1/c_0)\ln\sqrt{\det{KK_0^{-1}}} + O(1/c_0^2). \label{eq:F}
\end{equation}
The constant operator $K_{0\alpha\beta}=-\nabla^2\delta_{\alpha\beta}$ is introduced merely as
a convenient way of normalizing the contribution of fluctuations to $\mathcal{F}$. Thus, we are led 
to solve the spectral problem
\begin{equation}
K_{\alpha\beta}\xi_\beta = \lambda\xi_\alpha.
\end{equation}
In the terminology of Quantum Field Theory, the first term of~(\ref{eq:F}) is called 
the \textit{tree level} and the $1/c_0^n$ term the $n$-\textit{loop} order.
If $K$ is not positive definite the stationary point is unstable and the above expansion does 
not exists. The 1-loop term diverges in the continuum limit due to the short-distance fluctuations
and a short-distance cut-off is necessary.
In solid state physics it is naturally provided by the crystal lattice. In the numerical
computations we introduced the cut-off by discretizing $K$ with a step size $q_0dx=0.15$,
appropriate for MnSi. 
Thus, the fluctuation free energy is dominated by the short-distance fluctuations and 
depends strongly on the cut-off \cite{Muehlbauer09}.
Hence, the comparison of free energies of states computed with different cut-off schemes 
(different lattice discretization) is not meaningful. The low lying spectrum of $K$, however,
is well defined in the continuum limit and shows a weak depence on the cut-off.

The 1-loop approximation is valid if the terms of order $\xi^3$ and higher that are 
neglected in (\ref{eq:F}) do not give a large contribution. Since the leading contribution 
of the cubic term vanishes by symmetry, the contribution of the higher order
terms relative to the quadratic terms can be estimated by the ratio 
$\langle \xi^4\rangle/\langle\xi^2\rangle\sim\langle\xi^2\rangle=(1/c_0)\mathrm{Tr} K^{-1}/q_0V$.

In all the cases considered in this work, the $K$ operator has the generic form
\begin{equation} 
K = (-\nabla^2+U_\mathrm{S})I + U_\mathrm{A}\sigma_z + (E_\mathrm{T}+E_\mathrm{L})\sigma_y,
\label{eq:genK}
\end{equation}
where $I$ is the 2$\times$2 identity matrix, $\vec{\sigma}$ are the Pauli matrices,
$U_\mathrm{S}$ and $U_\mathrm{A}$ are functions of the coordinates, and $E_\mathrm{T}$ and
$E_\mathrm{L}$ differential operators linear in the derivatives.

\section{FM state}

The FM state is always a stationary point, with $\theta=0$ and $\psi$ undetermined
(may be taken as $\psi=0$). 
Its $K$ operator, 
\begin{equation}
K_{\alpha\beta} = (-\nabla^2+q_0^2h)\delta_{\alpha\beta} - 2q_0\partial_z\epsilon_{\alpha\beta},
\end{equation}
is readily diagonalized by Fourier transform, and its spectrum reads
\begin{equation}
\lambda_\pm = k_x^2+k_y^2+(k_z\pm q_0)^2 + q_0^2(h-1).
\end{equation}
where $\vec{k}$ is the wave vector of the eigenfunction.
The lowest eigenvalue is attained for $k_x=k_y=0$ and $k_z=\pm q_0$ and reads 
$\lambda_{\mathrm{min}}=(h-1)q_0^2$. Therefore, the FM state is
stable for $h>1$ and unstable for $h<1$.

\section{Conical helix}

\begin{figure}[t!]
\centering
\includegraphics[width=\linewidth,angle=0]{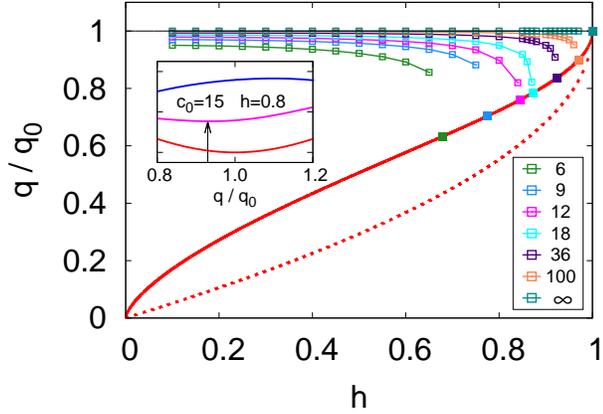}
\caption{The equilibrium wave number of the CH as a function of $h$ for the values
of $c_0$ indicated in the legend. The CH is a stationary point only on the left hand 
side of the broken red line, and its $K$ operator is positive definite only on the 
left hand side of the solid red line.
The inset displays the free energy to 1-loop level (pink) and its separate
tree level (red) and 1-loop (blue) contributions for $h=0.8$ and $c_0=15$.
The arrow signals the free energy minimum to 1-loop order.
\label{fig:q}}
\end{figure}

With the magnetic field directed along the $\hat{z}$ axis,
this stationary point has the form $\theta=\theta_0$ and $\psi=qz$, where $\theta_0$ and $q$
are constants.
The Euler--Lagrange equations are satisfied if and only if it holds the relation 
\begin{equation}
\cos\theta_0=\frac{h}{1-\Delta^2(q)}, 
\end{equation}
where
\begin{equation}
\Delta(q)=q/q_0-1. \label{eq:Delta}
\end{equation}
Since $|\cos\theta_0|\leq 1$, this stationary point exists only for $|\Delta|\leq\sqrt{1-h}$.
In Fig.~\ref{fig:q} the region of existence of the CH stationary point in the plane $(h,q)$ is limited by
the broken red line.
The value of $q$ is determined by minimizing the free energy in the region where the stationary point
is stable.

The fluctuation operator has the form of Eq.~(\ref{eq:genK}), with $U_\mathrm{S}=U_\mathrm{A}=q_0^2A/2$, 
where
\begin{equation} 
A = 1-\Delta^2 - h^2/(1-\Delta^2) \\
\end{equation}
is a constant and
\begin{eqnarray}
E_\mathrm{T} &=& -\mathrm{i}2q_0\sin\theta_0(\cos{qz}\partial_x+\sin{qz}\partial_y), \\
E_\mathrm{L} &=& \mathrm{i}2q_0\Delta\cos\theta_0\partial_z.
\end{eqnarray}
Notice that $A$ is positive for small $\Delta$ and $h<1$.

Due to the periodicity of the CH, the eigenfunctions of $K$ have the form
\begin{equation}
\xi_\alpha(\vec{r}) = \mathrm{e}^{\mathrm{i}\vec{k}\cdot\vec{r}}\eta_\alpha(z),
\end{equation}
where $k_x$ and $k_y$ are limited by the cut-off, $\pm\pi/a$,
$k_z\in [-q/2,q/2]$, and $\eta_\alpha(z)$ is periodic, with period $2\pi/q$.
The reduced spectral problem for $\eta_\alpha$ reads 
$\tilde{K}_{\alpha\beta}\eta_\beta = \lambda\eta_\alpha$,
with
\begin{equation} 
\tilde{K} = \left(-\partial_z^2+k_\mathrm{T}^2+q_0^2\frac{A}{2}\right)I + q_0^2\frac{A}{2}\sigma_z 
+ 2q_0\tilde{D}\sigma_y,
\end{equation}
where $k_\mathrm{T}^2=k_x^2+k_y^2$ and
\begin{equation}
\tilde{D} = \sin\theta_0(k_x\cos{qz}+k_y\sin{qz})+\mathrm{i}\Delta\cos\theta_0\partial_z,
\end{equation}
with periodic boundary conditions (BC) in $[0,2\pi/q]$,
This reduced spectral problem is solved numerically.

\begin{figure}[t!]
\centering
\includegraphics[width=\linewidth,angle=0]{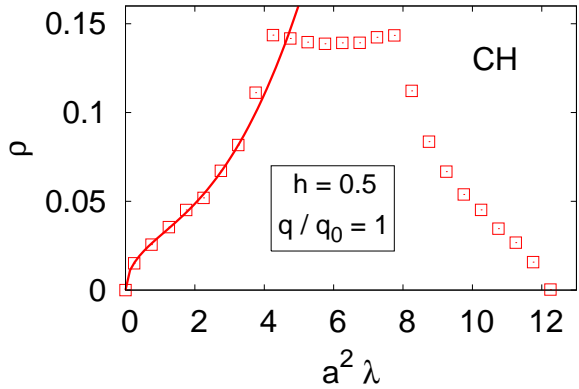}
\caption{Spectral density of the CH for the parameters displayed in the legend.
The line is a fit to the function $\rho(\lambda)=\sqrt{\lambda}(a_0+a_1\lambda+a_2\lambda^2)$
for $\lambda<2.5$.
\label{fig:chdensity}}
\end{figure}

The spectrum contains a zero mode corresponding to a Goldstone boson associated to the
global rotation of the helix about the magnetic field direction, which costs no energy. 
The spectral density,
\begin{equation}
\rho(\lambda) = \frac{1}{2V} \sum_i \delta(\lambda-\lambda_i),
\end{equation}
where $V$ is the volume, vanishes as $\sqrt{\lambda}$ when $\lambda\rightarrow 0$, and therefore the
fluctuation free energy is integrable and well defined. 
The spectral density is displayed in a typical case in Fig.~\ref{fig:chdensity}.

For $k_x=k_y=0$ the spectral problem can be analytically solved and gives two branches
\begin{equation}
\lambda_\pm = k_z^2 + 
\frac{q_0^2}{2} \left[ A \pm \left(A^2+16\Delta^2\cos^2\theta_0k_z^2/q_0^2\right)^{1/2} \right].
\end{equation} 
The $\lambda_-$ branch is the Goldstone mode while the $\lambda_+$ mode has a gap, $A$,
which vanishes on the boundary $q_\pm=q_0(1\pm\sqrt{1-h})$, 
where the stationary point ceases to exist.

The presence of the Goldstone modes does not invalidate the saddle point expansion, since 
the interactions of the Goldstone modes vanish at zero momentum, so that the
contribution of the zero mode to $\langle\xi^4\rangle$ vanish.
The the validity of the 1-loop approximation
is controlled by the gap, $A$, and the requirement is that
$1/c_0A$ is not large. In practice we require $1/c_0A<0.2$.

The Goldstone branch ($\lambda_-$) develops an instability if \mbox{$|q-q_0|$} is sufficently large: 
$\lambda_-$ becomes negative when $A<4\Delta^2\cos^2\theta_0$. The red solid line of 
Fig.~\ref{fig:q} signals the instability. The operator $K$ is positive definite
on the left hand side of the solid red line, and has negative eigenvalues in the
region between the solid and broken red lines.

The equilibrium wave number, $q_\mathrm{m}$, which depends on $c_0$ and $h$,
is determined by minimizing the free energy, 
which is plotted \textit{vs}. $q$ in the inset of Fig.~\ref{fig:q},
for a typical case to 1-loop level, 
Its separate tree level and 1-loop contributions are also shown.
Fig.~\ref{fig:q} displays $q_\mathrm{m}$ \textit{vs}. $h$ for fixed values of $c_0$.
At tree level ($c_0=\infty$) we have $q_\mathrm{m}=q_0$, independent of $h$. 
The 1-loop contribution shifts the free energy minimum to $q_\mathrm{m}<q_0$
(Fig.~\ref{fig:q}, inset). By incresing $h$ keeping $c_0$ constant, $q_m$ decreases
and approaches the instability line, which is however not continuously attained.
At a critical $h$ the minimum jumps discontinuously to a point on the 
instability line, marked with filled squares in Fig.~\ref{fig:q}.
This is a first order transition from the CH to another state which may be
the SKL, or an unknown state.
At tree level, however, the instability point is continuously attained and
there is a continuous transition to the FM state.
The critical field as a function of $T/T_0$ is represented by the red line 
in the phase diagram of Fig.~\ref{fig:phd}.

\section{Conical helicoid}

This stationary point is a one dimensional modulated structure that propagates in a direction that forms an
angle $\alpha$ with the magnetic field. If the propagation direction is along $\hat{z}$
and the magnetic field has components along $\hat{x}$ and $\hat{z}$,
the conical helicoid is described by two functions, $\theta(z)$ and $\psi(z)$, that where obtained in 
Refs.~\onlinecite{Laliena16a,Laliena16b,Laliena17}. It is characterized by two parameters,
the angle $\alpha$ and the period, $L$. The CH is recovered in the $\alpha\rightarrow 0$ limit,
while in the limiting case of $h_z=0$ we have $\theta=0$ and 
$\cos(\psi/2)=\mathrm{sn}(\sqrt{h_x}q_0z/\kappa)$, 
where $\mathrm{sn}(x)$ is the Jacobian elliptic function and
$\kappa$ the ellipticity modulus \cite{Dzyal64}.

\begin{figure}[t!]
\centering
\includegraphics[width=\linewidth,angle=0]{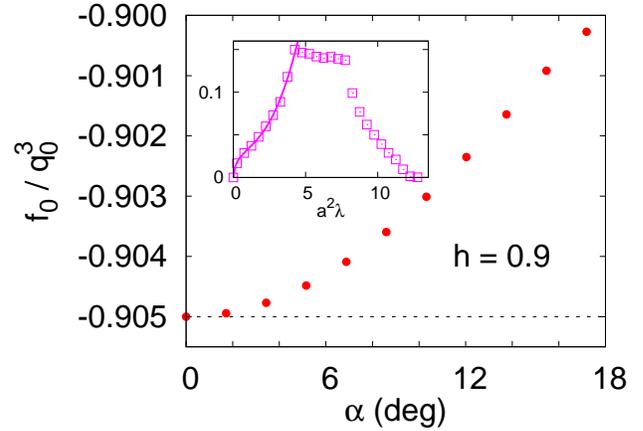}
\caption{Free energy density at tree level for the conical helicoid,
as a function of the angle, $\alpha$, between the magnetic field and the modulation
propagation direction. The inset shows the spectral density in a typical case ($h=0.9$,
$\alpha=15.5^o$, and $L/L_0=1.2$).
\label{fig:enerHL}}
\end{figure}

The $K$ operator is given by Eq.~(\ref{eq:genK}) with
\begin{eqnarray}
U_\mathrm{S} &=& q_0^2\frac{\sin^2\theta}{2}[\Delta^2(\psi^\prime)-1] \nonumber \\
&& \qquad\qquad + q_0^2(h_x\sin\theta\cos\psi +  h_z\cos\theta), \\
U_\mathrm{A} &=& q_0^2\frac{1}{2}(1-3\cos^2\theta)[\Delta^2(\psi^\prime)-1],  \\
E_\mathrm{T} &=& -\mathrm{i} 2q_0 \sin\theta(\cos{\psi}\partial_x+\sin{\psi}\partial_y), \\
E_\mathrm{L} &=& \mathrm{i}2(\psi^\prime-q_0)\cos\theta\partial_z,
\end{eqnarray}
where $\Delta(\psi^\prime)$ is given by Eq.~(\ref{eq:Delta}), substituting $q$ by $\psi^\prime$.
As in the case of the CH, $K$ is diagonalized with the help of the Fourier transform in
$x$ and $y$ and the Bloch-Floque theorem, remaining a spectral equation for the $z$ dependence
defined in an interval $z\in[0,L]$, with periodic BC. 
This reduced spectral equation is solved numerically. The spectral density in a typical case is
shown in the inset of Fig.~\ref{fig:enerHL}.

The spectrum of the conical helicoid contains one Goldstone boson, corresponding to 
the spontaneously broken translational symmetry along the direction of the modulation
propagation. Again, the spectral density
vanishes as $\sqrt{\lambda}$ at the origin (inset of Fig~\ref{fig:enerHL}) and the 1-loop 
contribution to the free energy is integrable and well defined. 
As discuss before, the Goldstone boson does not invalidate the saddle point expansion.

\begin{figure}[t!]
\centering
\includegraphics[width=\linewidth,angle=0]{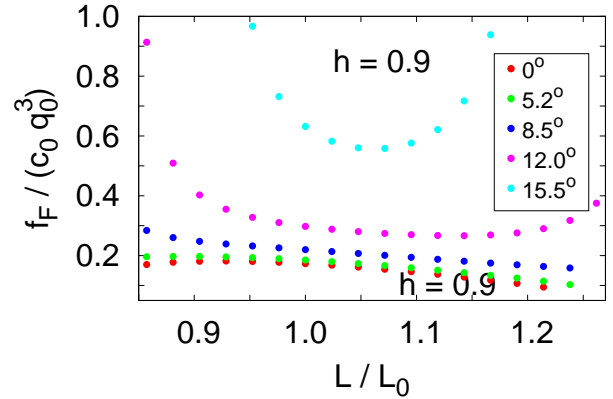}
\caption{One-loop contribution to the free energy density for the conical helicoid,
as a function of the period $L$ for the values of the angle between the magnetic field and the modulation
propagation direction displayed in the legend. 
\label{fig:enerFlucHL}}
\end{figure}

The period of the conical helicoid, $L$, is shifted from its tree level value, given in 
Ref.~\onlinecite{Laliena16a}, 
by the thermal fluctuations, analogously to what happens in the CH case. 
The conical helicoid has always higher free energy, at tree as well as at 1-loop level, than the CH.
That is, the free energy is always minimized by $\alpha=0$ for any $h$ and $c_0$.
Figs.~\ref{fig:enerHL} and~\ref{fig:enerFlucHL} display in a typical case how the tree level and
the 1-loop contribution to the free energy density increases with $\alpha$.
Hence, the conical helicoid is always an unstable stationary point, even though its $K$ operator 
is positive definite.

\section{Skyrmion lattice} 

\subsection{Circular cell approximation}

In what follows we take the magnetic field along the $\hat{z}$ direction: $\vec{h}=h\hat{z}$.
Consider an hexagonal SKL with lattice cells that contain a skyrmion core at its 
center \cite{Bogdanov94b}.
Each point of space, $\vec{x}$, belongs to a lattice cell and is parametrized by the 
coordinates of the cell center,
$\vec{r}_l$, where $l=1,2,3,\ldots$ labels the cells,
and the coordinates relative to the center cell, $\vec{r}$, so that
$\vec{x}=\vec{r}_l+\vec{r}$. For the derivatives we have obviously $\nabla_{\vec{x}}=\nabla_{\vec{r}}$.
It is convenient to use cylindrical coordinates $(r,\varphi,z)$ relative to the cell center,
instead of the cartessian coordinates $\vec{r}$. In the circular cell approximation
the stationary point within each cell is approximated by the axisymmetric solution
$\theta=\theta(r)$, $\psi=\pi/2+\varphi$, 
where $\theta(r)$ is the solution of the boundary value problem
\begin{equation}
\theta^{\prime\prime}+\frac{\theta^\prime}{r}-\frac{\sin\theta\cos\theta}{r^2}
+2q_0\frac{\sin^2\theta}{r}-q_0^2h\sin\theta = 0,
\end{equation}
and the prime stands for the derivative with respect to $r$.
The boundary conditions are $\theta(0)=\pi$ and $\theta(R)=0$, where $R$ is the radius of
the cylinder inscribed in the hexagonal cell \cite{Bogdanov94a}.

In the circular cell approximation the $K$ operator within each cell has the form of
Eq.~(\ref{eq:genK}), with
\begin{eqnarray}
U_\mathrm{S} &=& \frac{1}{r^2} - \frac{3\sin^2\theta}{2r^2} 
-\frac{\theta^\prime}{2}(\theta^\prime+2q_0) - q_0\frac{3\sin(2\theta)}{2r} \nonumber \\
&+& q_0^2h\cos\theta, \\
U_\mathrm{A} &=&  -\frac{\sin^2\theta}{2r^2} + \frac{\theta^\prime}{2}(\theta^\prime+2q_0) 
- q_0\frac{\sin(2\theta)}{2r}, \\
E_\mathrm{T} &=& \mathrm{i}2\left(\frac{\cos\theta}{r^2}-q_0\frac{\sin\theta}{r}\right)\partial_\varphi, \\
E_\mathrm{L} &=& -\mathrm{i}2q_0\cos\theta\partial_z.
\end{eqnarray}
This operator has been studied for isolated skyrmions ($R\rightarrow\infty$) on a plane
(eigenfunctions independent of $z$) in Ref.~\onlinecite{Schutte14}.

The lattice periodicity imply that the eigenstates of $K$ have the form
\begin{equation}
\xi_\alpha(\vec{x})=\mathrm{e}^{\mathrm{i}\vec{k}_\mathrm{T}\cdot\vec{x}}
\mathrm{e}^{\mathrm{i}k_zz}\Phi_\alpha(x,y), 
\end{equation}
where $\vec{k}_\mathrm{T} = k_x\hat{x}+k_y\hat{y}$ belong to the first Brillouin zone of the 
2D hexagonal reciprocal lattice, 
$k_z$ is limited by the short distance cut-off ($|k_z|<\pi/a$), 
and $\Phi_\alpha$ has the lattice periodicity. 
The reduced spectral problem becomes $\tilde{K}_{\alpha\beta}\Phi_\beta=\lambda\Phi_\alpha$, with
\begin{eqnarray}
\tilde{K}_{11} &=& -\nabla_\mathrm{T}^2 - 2\mathrm{i}\vec{k}_\mathrm{T}\cdot\nabla_\mathrm{T} 
+ U_\mathrm{S} + U_\mathrm{A}, \\
\tilde{K}_{22} &=& -\nabla_\mathrm{T}^2 - \mathrm{i}2\vec{k}_\mathrm{T}\cdot\nabla_\mathrm{T} 
+ U_\mathrm{S} - U_\mathrm{A}, \\
\tilde{K}_{12} &=& 2\left(\frac{\cos\theta}{r^2}-q_0\frac{\sin\theta}{r}\right) \left(\partial_\varphi 
+ \mathrm{i}\frac{\hat{z}\cdot(\vec{k}_\mathrm{T}\times\vec{r})}{r}\right)
\nonumber \\
&-& \mathrm{i}2q_0k_z\cos\theta,
\end{eqnarray}
and, $\tilde{K}_{21}=-\tilde{K}_{12}$. In the above equations we used the notation
$\nabla_\mathrm{T}=\hat{x}\partial_x+\hat{y}\partial_y$. 

Since $|\vec{k}_\mathrm{T}|\lesssim \pi/2R$, where $2R$ is the SKL cell diameter, 
they can be neglected in comparison with $k_z$ and
$\nabla_\mathrm{T}\Phi_\alpha$, which are of order $2\pi/a$ for modes rapidly varying at short distances, 
which provide the main contribution to the free energy. Given that $2R\gtrsim L_0$, the relative
contribution of the neglected terms is of order $a/L_0$, which is about 0.02 for MnSi.
 
With $\vec{k}_\mathrm{T}=0$, and ignoring the cell boundary effects,
the operator $\tilde{K}$ has cylindrical symmetry, what allows to reduce further the spectral 
problem by using the Fourier transform in $\varphi$, 
\begin{equation}
\Phi^{(n)}_\alpha(r,\varphi) = \mathrm{e}^{\mathrm{i}n\varphi}\phi_{\alpha,n}(r), \label{eq:Phi}
\end{equation}
so that it remains a system of radial equations for $\phi_{\alpha,n}(r)$ and $\lambda$. 

The number of circular Fourier modes is limited by a cut-off, $|n|\leq n_{\mathrm{max}}$, 
that increases with $R$, since we have to keep a constant short distance cut-off. 
This $n_{\mathrm{max}}$ is chosen so that the total number of modes equals the number of modes 
of a square lattice with a unit cell of the same area as the circular cell. 
If $a$ is the lattice parameter of the square lattice, the total number of modes is
$\pi R^2/a^2$. Then, if we take $dr=a$ for the step size of the radial discretization,
we get $n_{\mathrm{max}}=(\pi R/a-1)/2$. 

Imposing the periodicity to the function~(\ref{eq:Phi}) is subtle. 
The best approximation to a periodic function in the circular cell approximation is to 
identify opposite points on the circular boundary, that is  
$\Phi_{\alpha,n}(R,\varphi)=\Phi_{\alpha,n}(R,\varphi+\pi)$, what amounts
to $\phi_{\alpha,n}(R)=0$ for odd $n$ and no condition for even $n$.
We therefore set open (free) boundary conditions $\partial_r\phi_{\alpha,n}(R)=0$ 
for even $n$, what means that the fluctuations are extremal on the cell boundary.
Anyway, the fact that the fluctuation free energy is dominated by the short-distance
fluctuations means that there is little sensitivity to the boundary conditions.

It is clear that the circular cell approximation becomes exact in the large cell limit, 
$2R/L_0\rightarrow\infty$.
In this limit the neglected lattice momenta, $\vec{k}_\mathrm{T}$, vanish, the lattice cell 
consists on an axisymmetric central core surronded by a large FM background, and the sensitivity 
of the spectral problem to the BC disappears.
The results point out that it is good for $2R/L_0 \gtrsim 1$.

\begin{figure}[t!]
\centering
\includegraphics[width=\linewidth,angle=0]{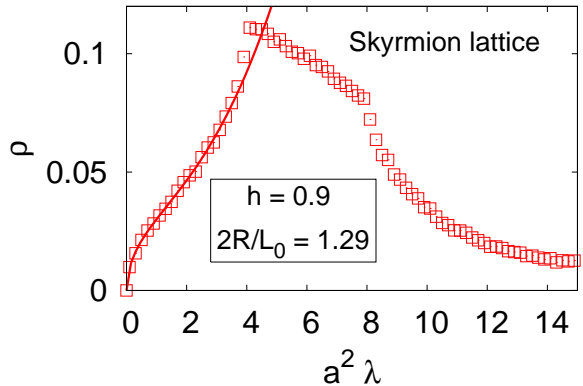}
\caption{Spectral density of the SKL for the parameters displayed in the legend.
The line is a fit to the function $\rho(\lambda)=\sqrt{\lambda}(a_0+a_1\lambda+a_2\lambda^2)$
for $\lambda<3$.
\label{fig:skldensity}}
\end{figure}

To deal with the radial spectral problem, it is convenient to work with
the reduced radial function $\chi_\alpha$, defined as
$\phi_{\alpha,n}(r) = \chi_{\alpha,n}(r)/\sqrt{r}$,
so that the spectral problem reads 
$A_{\alpha\beta}\chi_{\beta,n}=\lambda\chi_{\alpha,n}$, with
\begin{eqnarray}
A_{11} &=& -\partial_r^2 + \frac{n^2-1/4}{r^2} + k_z^2 + U_\mathrm{S} + U_\mathrm{A}, \\
A_{22} &=& -\partial_r^2 + \frac{n^2-1/4}{r^2} + k_z^2 + U_\mathrm{S} - U_\mathrm{A}, \\
A_{12} &=& \mathrm{i}n2\left(\frac{\cos\theta}{r^2}-q_0\frac{\sin\theta}{r}\right)
-\mathrm{i}q_0k_z\cos\theta,
\end{eqnarray}
and $A_{21}=-A_{12}$. The BC are $\chi_{\alpha,n}(0)=0$, since $\xi_\alpha$ has to be finite
at $r=0$, and
\begin{eqnarray}
&&\chi_{\alpha,n}(R) = 0 \qquad \qquad \qquad \quad n \mathrm{\ odd,} \\
&&\partial_r\chi_{\alpha,n}(R) = \frac{1}{2}\frac{\chi_{\alpha,n}(R)}{R} 
\qquad n \mathrm{\ even.}
\end{eqnarray}
The last equation results from the condition $\partial_r\phi_{\alpha,n}(R)=0$ for $n$ even.

The full spectrum of $A_{\alpha\beta}$ for each $n$ and different values of $R$
is obtained numerically with the help of the ARPACK software package \cite{ARPACK}.
The spectral density is displayed in Fig.~\ref{fig:skldensity} for a typical case.
From the spectrum, the 1-lopp contribution to the free energy is readily obtained 
as a function of $R$.

\subsection{Short distance approximation}

Since the short distance fluctuations dominate the fluctuation free energy, this can be
approximately obtained in a semi-analytic way. For short distance fluctuations the higher 
derivative terms entering $K$ are larger than the remaining terms, and thus it makes sense 
to split $K$ as $K = K^{(0)} +  Q$, where
\begin{equation}
K^{(0)}_{\alpha\beta} = (-\nabla^2 + 1/r^2)\delta_{\alpha\beta} - 2/r^2\partial_\varphi\epsilon_{\alpha\beta}.
\end{equation}
This is the (minus) Laplace operator that acts on $\vec{\xi} = (\xi_1,\xi_2)^{\mathrm{T}}$, which
are the coordinates on the spin tangent space in a local basis. 
The $1/r^2$ terms correspond to the connection associated to the local frame. 
The local rotation $\vec{\xi} = U\vec{\eta}$, with 
\begin{equation}
U = \left(
\begin{array}{rr}
\cos\varphi & -\sin\varphi  \\
\sin\varphi & \cos\varphi  \\
\end{array} 
\right)
\end{equation}
restores the global frame and the spectral equation $K^{(0)}\vec{\xi}=\lambda\vec{\xi}$ becomes
simply $-\nabla^2_{\vec{x}}\vec{\eta} = \lambda\vec{\eta}$. The solutions are plane waves
\begin{equation}
\vec{\eta}^{(i)}=\frac{1}{\sqrt{V}}\vec{u}^{(i)}\exp{(\mathrm{i}\vec{k}\cdot\vec{x})},
\end{equation}
with two polarizations, $i=1,2$, that can be chosen as
$\vec{u}^{(1)}=(1,0)^{\mathrm{T}}$, $\vec{u}^{(2)}=(0,1)^{\mathrm{T}}$, 
and eigenvalues $\lambda=k^2$. The volume can be written as $V=N_cV_cL_z$,
where $N_c$ is the number of skyrmion cells,
$V_c$ is the area of the unit cell in the XY plane and $L_z$ is the length
of the system in the $\hat{z}$ direction.
Hence, the eigenfunctions of $K^{(0)}$ are 
\begin{equation}
\vec{\xi}^{(i)}=\frac{1}{\sqrt{V}}U\vec{u}^{(i)}\exp{(\mathrm{i}\vec{k}\cdot\vec{x})}.
\label{eq:evecK0}
\end{equation}
It is important to bear in mind that the coordinates of any point can be expressed as
$\vec{x}=\vec{r}_l+\vec{r}$, where $\vec{r}_l$ are the coordinates of the center of the
cell to which the point belongs and $\vec{r}$ are the coordinates relative to the cell center.

\begin{figure}[t!]
\centering
\includegraphics[width=\linewidth,angle=0]{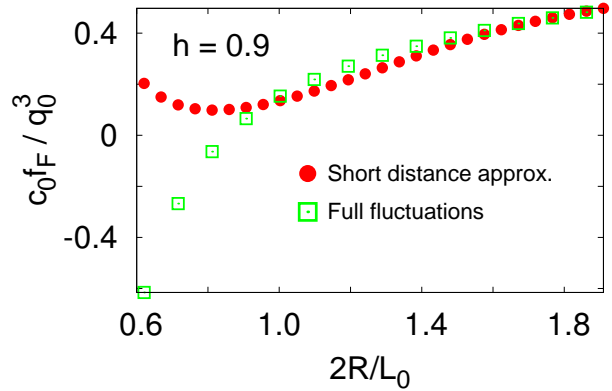}%
\caption{Free energy of fluctuations at one loop order in the short distance
approximations (red circles) and taking into account the full set of fluctuations
(open green squares).
\label{fig:sda}}%
\end{figure}

The fluctuation energy to one loop order is given by
\begin{equation}
c_0f_\mathrm{F} = \frac{1}{2V}\left[\mathrm{Tr}\ln(K^{(0)}+Q)-\mathrm{Tr}\ln K^{(0)}\right].
\end{equation}
For short distance fluctuations $K^{(0)}$ is much larger than $Q$ and the fluctuation free
energy can be approximated by
\begin{equation}
c_0f_\mathrm{F} = \frac{1}{2V}\mathrm{Tr} (QK^{(0)\,-1}).
\end{equation}
Using the basis of eigenfunctions of $K^{(0)}$, given by Eq.~(\ref{eq:evecK0}), the fluctuation
free energy in the short distance approximation reads
\begin{equation}
c_0f_\mathrm{F} = \int \frac{d^3k}{(2\pi)^3}\frac{1}{k^2}\sum_{i=1}^{2}
\left\langle\left.\vec{\xi}^{(i)}\right|Q\left|\vec{\xi}^{(i)}\right.\right\rangle,
\label{eq:intk}
\end{equation}
where
\begin{equation}
\left\langle\left.\vec{\xi}^{(i)}\right|Q\left|\vec{\xi}^{(i)}\right.\right\rangle =
\frac{1}{V}\sum_l \int dz \int_{V_\mathrm{c}}d^2r\,
\mathrm{e}^{-\mathrm{i}\vec{k}\vec{x}}\vec{u}_i\cdot
U^{\dagger}QU\vec{u}_i \mathrm{e}^{\mathrm{i}\vec{k}\vec{x}}.
\end{equation}
The integrand in the above expression turns out to be independent of $\vec{r}_l$ and $z$,
and therefore
\begin{equation}
\left\langle\left.\vec{\xi}^{(i)}\right|Q\left|\vec{\xi}^{(i)}\right.\right\rangle =
\frac{1}{V_\mathrm{c}}\int_{V_\mathrm{c}}d^2r\,
\mathrm{e}^{-\mathrm{i}\vec{k}\vec{r}}\vec{u}_i\cdot
U^{\dagger}QU\vec{u}_i \mathrm{e}^{\mathrm{i}\vec{k}\vec{r}}.
\end{equation}
Some algebraic manipulations allow to write the above expression as
\begin{equation}
\sum_{i=1}^2\left\langle\left.\vec{\xi}^{(i)}\right|Q\left|\vec{\xi}^{(i)}\right.\right\rangle =
\frac{2}{R^2}\int_0^R dr \, r \Upsilon(r),
\label{eq:matel}
\end{equation}
where
\begin{eqnarray}
\Upsilon(r) &=& -3\frac{\sin^2\theta}{r^2} - 3q_0\frac{\sin(2\theta)}{r}-\theta^{\prime\,2}-2q_0\theta^\prime
+ 2q_0^2h\cos\theta \nonumber \\
&+& 4\left(\frac{1+\cos\theta}{r^2}-q_0\frac{\sin\theta}{r}\right).
\end{eqnarray}
The matrix elements~(\ref{eq:matel}) are independent of $\vec{k}$, and therefore the integral
over $\vec{k}$ in~(\ref{eq:intk}) can be readily performed. With a cutoff $\pi/a$ for $k$, we obtain
\begin{equation}
c_0f_\mathrm{F} = \frac{1}{2\pi R^2a}\int_0^Rdr\,r\Upsilon(r)
\end{equation}
in the short distance approximation.

Fig.~\ref{fig:sda} displays the fluctuation free energy in the short distance approximation
as a function of $2R/L_0$ for $h=0.9$. For comparison, the fluctuation free energy obtained
in the circular cell approximation taking into account the full set of fluctuations is also displayed.
Notice that the short distance approximation is very good for $2R/L_0>1$.
As a conservative criterium, we consider the circular cell approximation reliable if $2R/L_0>1.25$. 
The phase boundary that limits the phase diagram region where the SKL is (meta)stable 
computed in the short distance approximation coincides 
essentially with the phase boundary obtained taking into account the full set of fluctuations, 
which is displayed in Fig.~\ref{fig:phd}.

\subsection{Results}

\begin{figure}[t!]
\centering
\includegraphics[width=\linewidth,angle=0]{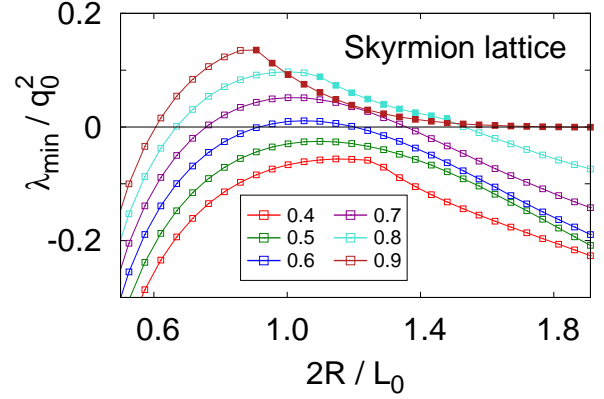}
\caption{The lowest eigenvalue of $K$ for the SKL as a function of the cell radius $R$
for the values of $h$ displayed in the legend. The filled squares correspond to 
Goldstone modes.
\label{fig:skleval}}
\end{figure}

The lowest eigenvalue ($\lambda_{\mathrm{min}}$) of $K$ as a function of $2R/L_0$ is 
displayed in Fig.~\ref{fig:skleval} for several values of $h$. 
In some cases there is level crossing, signaled by the cusps of the curves.
The negative eigenvalues correspond in all cases to modes with $k_z\neq 0$, that
propagate along the magnetic field direction. The exception are the SKL with
very small lattice diameter, which are unstable for all $h$,
in which case the lowest lying modes have $k_z=0$ and
$n=0$, and show the tendency of the lattice cell to expand.
The SKL cannot exists for $h <0.57$, as in this case
$\lambda_{\mathrm{min}}$ is negative for all $R$.
However, it is positive in a limited interval of $R$ if $0.57<h<1$,
and for sufficiently large $R$ if $h>1$.
The tree level free energy has no minimum in these intervals of $R$ and thus the
SKL is unstable for all $h$ at tree level ($c_0\rightarrow\infty$).
As we shall shown below, the 1-loop free energy may turn the SKL (meta)stable for $h>0.57$. 

\begin{figure}[t!]
\centering
\includegraphics[width=\linewidth,angle=0]{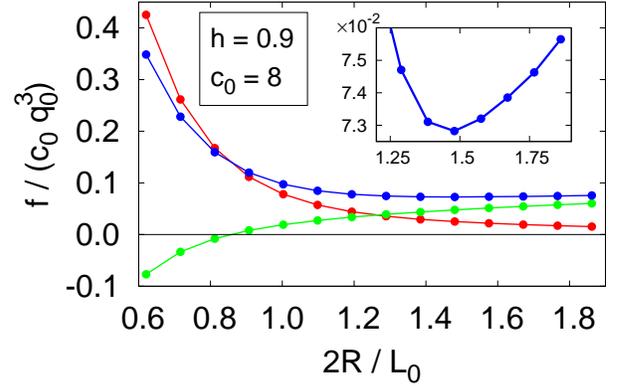}
\caption{The tree level (red) and 1-loop (green) contributions to the free energy density 
\textit{vs.} the cell radius for $h=0.9$ and $c_0=8$. In blue, the total free energy density. 
The vicinity of the minimum is magnified in the inset.
\label{fig:sklfe}}
\end{figure}

The SKL has zero modes corresponding to the Goldstone bosons associated to the
breaking of the continuous translational symmetry. Actually, the translational
symmetry is not continuous, since there is an underlying crystal lattice and
the would be zero modes acquire a gap of order $a/2R\approx 0.025$. Not surprisingly,
the circular cell approximation fails to reproduce the Goldstone modes for small $R$, 
but it reproduces them fairly well for large enough $R$. The lowest lying mode for $h=0.8$ and $0.9$
corresponds to the Goldstone boson in an interval of $R$. They are marked with filled
squares in Fig.~\ref{fig:skleval}. These modes are identified as Goldstone bosons since
the eigenfunctions have the quantum numbers of the translational zero modes of the
isolated skyrmions: $k_z=0$ and $n=\pm 1$.
For $2R/L_0>1.25$, when we consider the circular cell approximation valid, the Goldstone modes gap
is rather close to its expected value of $0.025 q_0^2$.

The tree level and 1-loop contributions to the free energy of the SKL, taking into account
the whole set of fluctuations,
are displayed as a function of $2R/L_0$ for $h=0.9$ and $c_0=8$ in Fig.~\ref{fig:sklfe}. 
At tree level the free energy has no local minimum and the SKL is unstable for all $h$.
However, the contribution of the fluctuations at 1-loop level produces a minimum of the
free energy at $2R/L_0\approx 1.5$. This behaviour is generic:
a local minimum appears for low enough $c_0$ if $h>0.57$,
and the SKL becomes at least metastable.
The phase diagram region where the SKL is (meta)stable is encircled by a magenta line 
in Fig.~\ref{fig:phd}.
In the light magneta region it is the equilibrium state.
We consider the computation reliable in the region filled with magenta stripes, 
where both the 1-loop approximation and the circular cell approximation are reliable as 
$\langle\xi^2\rangle<0.2$ and $2R/L_0>1.25$.
Similar results are obtained if the 1-loop free energy is computed by using the short
distance approximation.

\section{Phase diagram}

Let us discuss first the tree level, ignoring the fluctuations. The FM state is stable 
for $h>1$ and unstable for $h<1$, and the opposite happens with the CH. The conical helicoid 
has always higher free energy than its limiting case, the CH, and thus is unstable. 
The SKL is also unstable, since its free energy has no local minimum in the interval
of lattice sizes were $K$ is positive definite.
Thus, the tree level phase diagram is very simple: the equilibrium state is the CH 
for $h<1$ and the FM state for $h>1$, with no metastable state. 
The equilibrium wave number of the CH is $q_0$ for all $h$.

\begin{figure}[t!]
\centering
\includegraphics[width=\linewidth,angle=0]{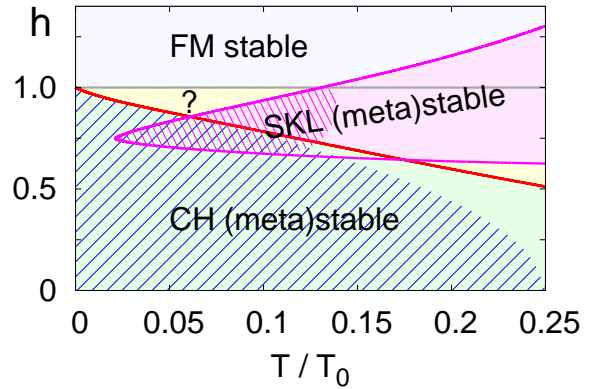}%
\caption{Phase diagram. The SKL is stable in the magenta region.
The computations for the CH and the SKL are reliable in the regions filled with blue 
and magenta stripes, respectively. None of the known stationary points are stable in the yellow regions. 
A new modulated state is expected in the region signaled with a question mark (?), where the
1-loop approximation is expected to be reliable.
\label{fig:phd}}%
\end{figure}

Thermal fluctuations change the scenario. 
The phase diagram to 1-loop order is displayed in Fig.~\ref{fig:phd}.
Due to wave number renormalization,  
the CH becomes unstable at a critical magnetic field that depends on $c_0=T/T_0$
(red line in Fig.~\ref{fig:phd}).
Since the critical field is smaller than one, the FM state is still unstable when
the CH becomes unstable, and therefore there is a region in the phase diagram, bounded by the
red and gray lines in Fig.~\ref{fig:phd}, where both the CH and the FM state are unstable. 
The SKL is at least metastable in the region encircled by the magenta line 
in Fig.~\ref{fig:phd}, and it is the equilibrium state in the light magenta region.
Where both states coexist, the CH has lower free energy than the SKL,
but the comparison is not meaningful, even though they are of the same order of magnitude, 
since the free energies have been computed with different cut-off schemes (a square
lattice for the CH and the circular cell approximation for the SKL).
To determine the equilibrium state in this region
we have to go beyond the circular cell approximation and perform an exact calculation 
of the 1-loop SKL free energy.
Finally, the conical helicoid is unstable against the tilting towards the CH 
everywhere in the phase diagram.

An intermediate region, colored in yellow in Fig.~\ref{fig:phd}, appears in the phase 
diagram in which none of the known stationary points (FM, conical helix, conical helicoid, 
and skyrmion lattice) are stable.
The equilibrium state in the intermediate region signaled with a question mark
(?) in Fig.~\ref{fig:phd}, where the saddle point expansion is expected to be valid,
will likely be described by an unknown stationay point with modulations in the three 
dimensions \cite{Rybakov13,Rybakov15,Leonov16}.

The computations for the CH and the SKL are estimated to be reliable in the regions
marked in Fig.~\ref{fig:phd} by the blue and magenta stripes, respectively.
Notice also that the present analysis is not valid at low magnetic field, 
where many stationary points of CH type are nearly degenerate and the theory of Brazovskii 
type is necessary \cite{Brazovskii75,Janoschek13}.

We may estimate the region of the phase diagram of MnSi where the present results apply.
For MnSi we have \mbox{$L_0=190$ \AA} and \mbox{$a=4.56$ \AA}, so that $q_0a=0.15$.
Mean field theory provides the relations $D/J = \tan q_0a = 0.15$, and 
\begin{equation}
\frac{k_\mathrm{B}T_\mathrm{c}}{JS^2} = \frac{2}{3}\left(2+\sqrt{1+D^2/J^2}\right) \approx 2
\end{equation} 
for the zero field critical temperature, $T_\mathrm{c}$. Therefore, we have the estimate
$T_0/T_\mathrm{c}\approx J/2D = 3.3$. The computations presented here
are reliable for $T/T_0\approx 0.15$ (Fig.~\ref{fig:phd}), that is for $T/T_\mathrm{c}\approx 0.5$.
Given that $\mbox{$T_\mathrm{c}=29.5$ K}$, the present results predict the appearance of
a skyrmion lattice and an intermediate unknown state separating the CH and forced FM phases 
in MnSi for \mbox{$T\lesssim 15$ K}. 

\section{Final remarks}

To conclude, we want to stress again that at low $T$ the CH and forced FM phases 
are not connected, but are separated by an SKL and an intermediate modulated phase
of unknown type. Thermal fluctuations are 
the crucial ingredient both to destabilize the CH state and to stabilize the SKL.
The low $T$ phase diagram of cubic helimagnets might therefore be richer than expected
and deserves a careful experimental investigation.

The idea that thermal fluctuations may stabilize the SKL was put forward in 
Ref.~\onlinecite{Muehlbauer09}, where a Landau--Ginzburg model with a strongly
fluctuationg modulus of the magnetic moment was studied.
The conclusion was that a SKL was formed in a small region of the phase diagram
that can be identified with the so called A phase of MnSi.
The computations were performed in Fourier space and thus the solitonic nature
of the SKL was not manifest. 
In the present calculations, valid at lower $T$, the fluctuations of the magnetic moment 
are small: its average modulus is given to 1-loop level by 
$|\langle\hat{n}\rangle|=1-\mathrm{Tr} K^{-1}/(2c_0q_0V)$,
and the results cannot be compared with those of Ref.~\onlinecite{Muehlbauer09}.
It would be interesting to study that model with the 
methods of this paper, in which the solitonic nature of the SKL is manifest.

\begin{acknowledgments}
The authors are greateful to A. N. Bogdanov and A. O. Leonov for interesting discussions. 
The authors acknowledge the Grant No. MAT2015-68200- C2-2-P from the Spanish Ministry of 
Economy and Competitiveness. 
This work was partially supported by the scientific JSPS 
Grant-in-Aid for Scientific Research (S) (Grant No. 25220803), 
and the MEXT program for promoting the enhancement of research universities, 
and JSPS Core-to-Core Program, A. Advanced Research Networks.
\end{acknowledgments}

\bibliography{references}

\end{document}